\documentclass[showpacs,floatfix,aps]{revtex4}
\usepackage{amssymb,amsmath}
\usepackage[dvips]{graphics,color}
\usepackage{epsfig}
\usepackage{float}
\usepackage[T1]{fontenc}
\usepackage{lscape}
\usepackage{cleveref}

\usepackage{lscape}
\newcommand{\bea}{\begin{array}}
\newcommand{\ear}{\end{array}}

\newcommand{\bege}{\begin{equation}}
\newcommand{\enge}{\end{equation}}

\newcommand{\beq}{\begin{eqnarray}}\newcommand{\benu}{\begin{enumerate}}\newcommand{\enu}{\end{enumerate}}
\newcommand{\eeq}{\end{eqnarray}}

\begin{document}

\title{Central accumulation of magnetic flux in massive Seyferts as possible engine to trigger UHECRs}

\author{R. C. Anjos}
\email{ritacassia@ufpr.br}
\author{C. H. Coimbra-Ara\'ujo}
\email{carlos.coimbra@ufpr.br}


\affiliation{Departamento de Engenharias e Ci\^encias Exatas,
  Universidade Federal do Paran\'a (UFPR),\\ 
Pioneiro, 2153, 85950-000 Palotina, PR, Brazil.}



\begin{abstract}

In the present paper we investigate the production of ultra high energy cosmic rays (UHECRs) from Seyferts. We discuss the UHECR luminosities obtained by two possible engine trigger models: pure radiative transfer and the energy extraction from poloidal magnetic flux. The first case is modeled by Kerr slim disk or Bondi accretion mechanisms. Since it is assumed that the broad-band spectra of Seyferts indicate that at least the outter portions of their accretion disks are cold and geometrically thin, and since our results point that the consequent radiative energy transfer is inefficient, we build the second approach based on massive Seyferts with sufficient central poloidal magnetic field to trigger an outflow of magnetically driven charged particles capable to explain the observed UHECRs and gamma-rays in Earth experiments from a given Seyfert source. 
\end{abstract}

\maketitle


\section{Introduction}

One of the current most familiar conundrums of particle astrophysics is surely the origin of ultra-high energy cosmic rays (UHECRs). UHECRs are most probably accelerated by astrophysical shocks as postulated, for instance, by Fermi \cite{fermi}. In this scenario charged particles could be accelerated by clouds of magnetised gas moving within our galaxy. He created this model partly to refute the arguments of Teller and Richtmeyer and of Alfv\'en on the idea that cosmic rays come dominantly from the Sun. Nevertheless it is well known that Fermi's original mechanism is too slow to be effective for UHECRs and nowadays it was adapted to a novel diffusive shock acceleration mechanism, as a way to accelerate charged particles efficiently at shock fronts. Other possible mechanisms are the external shock phase \cite{vietri}, unipolar inductors \cite{olinto1}, magnetic reconnection acceleration and re-acceleration in sheared jets \cite{olinto1}. A detailed sketch about all possible mechanisms and sources can be found in \cite{nagano,cronin,olinto1}.

Essentially, in the concept of diffusive shock accelerations (internal shocks), the particle must be confined within the acceleration region long enough to gain energy. Macroscopic motion is coherent and particles can gain energy as they bounce back and forth, making the energy gain $\Delta E / E \sim \beta$, where $\beta$ is the average velocity of the scattering centers in units of $c$ \cite{bell,blandford}. In a single-shot acceleration process, such as might occur, e.g., near a neutron star, a high voltage could be generated between two regions of the source. Suitable shocks are thought to occur near active galactic nuclei (AGNs), near black holes (BHs) and, effective at lower energies, in association with supernova remnants. Above $10^{20}$ eV there is a dearth of objects that satisfy the main aspects of this mechanism as first pointed out by Hillas \cite{hillas}: radio galaxies, colliding galaxies, AGNs, magnetars, gamma ray bursts (GRBs) and perhaps galactic clusters might host the right conditions. On the other hand, objects such as supernovae remnants and magnetic A stars are considered incapable of accelerating particles to this energy.

Assuming that the acceleration region must be of a size $R$ to match the Larmor radius , i.e. $r_L \leq R$, with $r_L = E/(ZeB)$ the Larmor radius of the particle of energy $E$ in the source magnetic field $B$, assuming the particle being accelerated and that the magnetic field within it must be sufficiently weak to limit synchrotron losses, it can be shown that the total magnetic energy in the source grows as $\Gamma^5$, where $\Gamma$ is the Lorentz factor of the particle. This analysis leads to the conclusion that putative cosmic ray sources might be strong radio emitters with radio powers  $\gg 10^{41}$ ergs s$^{-1}$, unless protons or heavier nuclei are being accelerated and electrons are not. Analyses such as that in \cite{kelner} show that this inequality is at least $> 10^{44}$ ergs s$^{-1}$ and amongst the few nearby sources that satisfy this limit are Centaurus A and M87. Interestingly, if acceleration takes place due to the internal shock mechanism, one may expect a strong neutrino signature due to proton interactions with the radiative background \cite{angelis,abdo}. 

On the other hand, non-blazar AGNs have came up as strong candidates to be $\gamma$-ray and possibly UHECR emitting sources, since observations with $\gamma$-ray telescopes can provide robust additional information to the origin of UHECRs. For example, the detection of Narrow Line Seyfert 1 galaxies in the Fermi-LAT energy regime reveals essential hints to develop the understanding of jet formation, radio-loudness and particle acceleration at non-blazar cores \cite{rieger1,rieger2}. It is estimated that only $\sim 1\%$--$5\%$ of galaxies contain bright Seyfert nuclei \cite{huchra1992,greene2007}. Nevertheless, in this regard, Seyfert are galaxies that comprise the most abundant class of AGN in the local Universe with strong emission lines, dominating the population of radio-quiet AGNs. This class of nearby galaxies are studied in many ways, specially by the nuclear activity point of view (see, e.g. \cite{komossa2007,ho2008}, for a review). Some authors refer Seyfert as a class of low-luminosity AGN (LLAGN) with moderately radio-loudness \cite{ho2002,gallimore2006} and with very high surface brightnesses whose spectra reveal strong, high-ionisation emission lines \cite{seyfert1943,kwan1981,mathur2000}. Other important aspects regarding Seyferts are the observation of the presence of high X-ray cores \cite{nandra1994,roberts2000,panessa2006}, inefficient central accretion mechanism that is translated in low accretion rates and Eddington-ratio sequence that extends down to $\lambda \sim 10^{-4}$ \cite{sikora2007,ho2009}, and, most important, at such lowest accretion rates, it is considered that an increasing fraction of the accretion energy is guided into a relativistic jet, making prominent the role of central poloidal magnetic fields and leading the emitted energy to be mainly kinetic rather than radiative \cite{livio1999,ho2008,marek}. 

The main models of cosmic ray acceleration in Seyfert Nuclei were proposed in \cite{tajima,kar,uryson} and LLAGNs, specially Seyfert galaxies nearby, were suggested as probable sources of UHECRs by \cite{kalm,ury,ammando}. Our present investigation endorses this idea, constraining that this occur mainly for massive Seyferts, comparing our results and upper limits to that obtained by the Fermi-LAT experiment. Since the broad-band spectra of Seyferts and other LLAGNs indicate that at least the outer portions of their accretion disks are cold and geometrically thin \cite{lawrence2005,nehmen2006,yu2010,nehmen2012}, we begin the article calculating the role of Kerr slim disk and spherical symmetry (Bondi) accretion rate around black holes with mass range of that found in Seyferts cores (Section \ref{sec:kerr}) to see how much radiative energy should be transfered to trigger UHECRs in Seyferts. The comparison of this with the luminosity of UHECRs from seven Seyfert sources calculated from FERMI-LAT integral $\gamma$-ray flux can give the conversion rate thresholds ($\eta_{pr}^{UL}$) of the AGN luminosity to trigger the acceleration of protons. Since it is expected that in Seyferts the accretion is inefficient and our results for Bondi accretion indicate great values of $\eta_{pr}^{UL}$ to accelerate protons (indicating that possibly the transfer of energy from Seyfert accretion radiative processes to accelerate protons is inefficient), in Section \ref{sec:mag} it will be considered an alternative mechanism by which a poloidal magnetic field attached to the slim disk and the central black hole can handle to extract energy from the disk or the black hole, in the form of a pure magnetically driven charged particle wind. It will be assumed that in Seyferts the role of central poloidal magnetic fields is proeminent \cite{ho2008,marek} allowing us to calculate the upper limits of the magnetic field and consequently the upper limits of energy extraction in Seyferts to proper trigger UHECRs. Comparing such upper limits to UHECRs upper limits permits us to say that at least massive Seyferts could have proper attributes to trigger UHECRs. Finally in Sections \ref{sec:discussion} and \ref{sec:conclusion} we have issued some discussion and concluding remarks.

\begin{figure}[h]
  \centering
  \includegraphics[width=12cm]{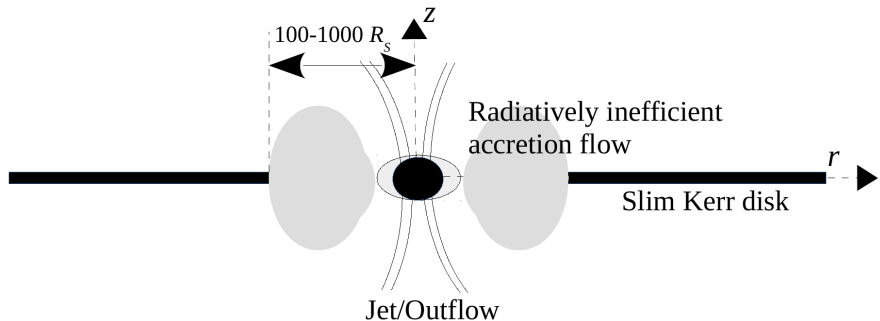}\\
\caption{Model of a Seyfert accretion disk, that is slim, cold and presenting an inefficient accretion flow due to the great distance from the central black hole. The role of the central poloidal magnetic field is to produce a jet of magnetically driven charged particles, triggering the emission of UHECRs. A approximately spherical accretion symmetry could be modeled for such system, privileging a Bondi accretion.}
\label{fig:sketch}
\end{figure}

\section{Radiative transfer and poloidal magnetic field energy extraction}\label{sec:kerr}

In previous works we studied upper limits of UHECR luminosities of ultra-high-energy cosmic rays from Seyferts and radio-galaxies. The method takes into account the calculation of upper limits of cosmic ray luminosities from the upper limit of the integral gamma ray flux measured by FERMI-LAT, VERITAS, MAGIC and HESS observatories \cite{fermi1,veritas, magic, ta, hess}. This limit was obtained for several sources, including radio-galaxies, active galactic nucles and Seyfert using propagation of UHECRs \cite{supa,vitor}. The method considers that UHECRs produce gamma rays as they propagate from the source to Earth. The gamma rays generated by such mechanism contribute to the total flux observed.

\begin{table}
\centering
\caption{Columns show the source name, the redshift, the mass source according to
  references~\cite{lisa,michael}, the mass black hole according to
  references~\cite{vasu,franca} and using the method of Reines~\cite{reines}, the upper limit on the proton UHECR
  luminosity, the 14 - 195 $keV$ luminosity according to reference~\cite{fermisey} and
  the X-rays luminosity calculated with equation \ref{equationlx}. }
\label{tab:1}
\begin{tabular}{|c|c|c|c|c|c|c|}
\hline \hline \textbf{Source name} & \textbf{$z_s$}  &
\textbf{\hbox{log}$M_{\odot}$}  & \textbf{\hbox{log}$M_{BH}$} &\textbf{$L_{CR}^{UL}$}&\textbf{\hbox{log}$L_{X}$}& \textbf{\hbox{log}$L_{X}^{Theory}$} \\
                            &                             &   &  
                            &[erg s$^{-1} \times 10^{45}$]&[erg s$^{-1}$] &[erg s$^{-1}$]\\ \hline
    NGC 985       & 0.04353 &10.6      &         8.39      &1.32 & 44.10&44.28\\ \hline
  NGC 1142      & 0.02916  & 10.7     &         8.53      &0.84  &44.23&44.22\\ \hline
  2MASX J07595347+2323241     &0.03064   & 10.57 &        8.34 &1.67&43.79&44.23   \\ \hline
   CGCG 420-015  & 0.02995 & 10.63      &         8.43      & 1.60&43.69&44.32\\ \hline
    MCG-01-24-012 &0.02136  & 10.16       &         7.77      &  1.18&43.59&43.63 \\ \hline
  LEDA 170194 & 0.04024 &  10.59      &         8.37      & 2.32&44.17&44.26 \\ \hline
   Mrk 520   & 0.02772  &  10.4      &         8.11      & 1.69&43.70&44.00 \\ \hline
 \end{tabular}
\end{table}

The production of the luminosity of cosmic rays obtained from massive Seyferts is also explicitly linked to mechanisms of black hole accretion \cite{ioana, biermann}. The consequent magnetic flux from rotating black holes (Blandford-Znajek mechanism) originated at the center of the source \cite{marek, mckinney}, can also produce a fraction of these luminosities to be converted into luminosity of cosmic rays ($\eta_{CR}$). These results indicate once again a massive Seyfert correlation as source of cosmic rays \cite{roldao, carlos}. The main characteristics of the sources are summarized in table~\ref{tab:1}.

The comparison between the magnetic luminosity and the UHECR luminosity can be seen in \cite{carlos}, where it is assumed a Bondi accretion mechanism with a central Kerr black hole. 

\subsection{Kerr slim disk and Bondi accretion in Seyferts}

Here it will be considered two kinds of accretion mechanism. Firstly, the spheric symmetric Bondi accretion model $\dot{M} = 4\pi c_S\rho_B r_B^2$ \cite{bondi}, where $\dot{M}$ is the accretion rate, $c_S$ is the sound speed in the medium, $r_B \approx GM/c_S^2$ is the Bondi accretion radius, and $\rho_B$ is the gas density at that radius, which produces, by friction and other radiative processes, considerable bolometric luminosities. Such parameters for nearby Seyfert cores are listed in \cite{ho2009}. Another possibility is to derive a slim disk in the Kerr spacetime, producing a most realistic accretion 

\begin{equation}
\dot{M} = -2\pi \Sigma \Delta^{1/2}\frac{V}{\sqrt{1-V^2}},
\end{equation}
\noindent
where $\Sigma = \int_{-h}^{+h}\rho dz$ is the disk surface density and V, defined by the relation $u^r=V\Delta^{1/2}/(r\sqrt{1-V^2})$, is the gas radial velocity as measured by an observer at fixed $r$ who corotates with the fluid. $\Delta$ is one of the standar Kerr metric coefficients, given by $\Delta = r^2 + \frac{a^2}{c^2} - 2\frac{GM}{c^2}r$, where $a$ is the black hole spin (see, e.g. \cite{sadowski} and \cite{fragile}).

\begin{figure}[h]
  \centering
  \includegraphics[width=12cm]{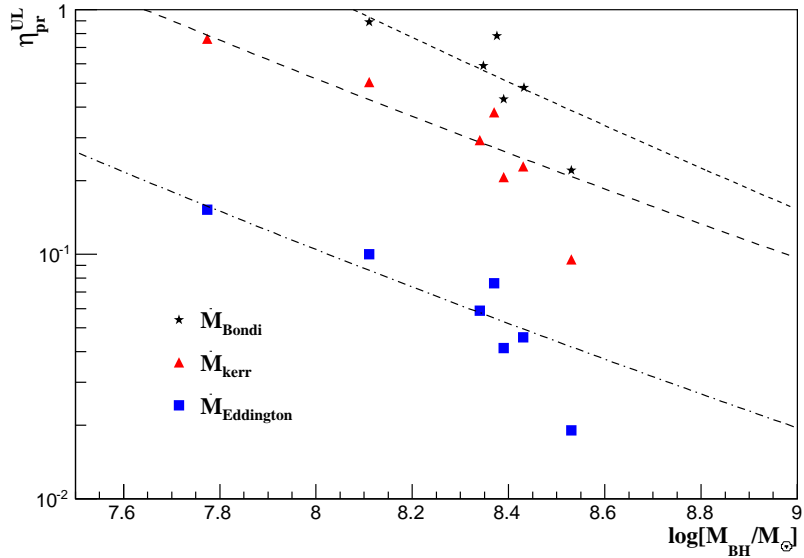}\\
\caption{Comparison between $\eta_{pr}^{UL}$ and $log ( M_{BH}/M_\odot )$
  using three different models of accretion: Bondi, Kerr slim disk and
  Eddington. (See the study of the stability of such systems at
  \cite{stability1,stability2}.) The $\eta_{pr}^{UL}$ is calculated
  using the method presented in \cite{roldao,carlos} neglecting the radiation losses during acceleration of
UHECR. The original method uses the mass of the galaxy to calculate the UHECR upper limit
  flux and here the model was adapted to use the black hole mass. For
  the Seyfert case, since the accretion is inefficient, $\eta_{pr}^{UL}$
  will have values between the Kerr slim model (triangles) and the Bondi
  model (stars). Note that at the Bondi threshold, the $\eta_{pr}^{UL}$
  values are larger than in the Kerr slim disk threshold case,
  indicating that the radiative system should transfer more energy to
  accelerate cosmic rays when the accretion mechanism is inefficient as
  that found in Seyferts and therefore only massive Seyferts $(>10^{8.2}
  {M_{\odot}})$ are sources capable of accelerating cosmic rays.  In the figure, each point corresponds to one of the Seyferts presented at Table \ref{tab:1}.}
\label{fig:fit}
\end{figure}

\noindent Associated to this, it is possible to use the upper limit $\eta_{pr}^{UL}$ conversion fraction of Kerr black hole accretion energy into energy to accelerate the UHECRs (protons), see the full description in \cite{roldao}: 

\begin{equation}
\eta_{pr}^{UL} = L_{pr}^{UL}/L_{acc}^{Theory}, 
\end{equation}

\noindent
with

\begin{equation}
L_{pr}^{UL} = \frac{4\pi D^{2}_s(1+z_s)\langle E \rangle_{0}}{ \ K_{\gamma} {\mathop{\displaystyle
\int_{E_{th}}^{\infty} dE_\gamma\ P_{\gamma}(E_{\gamma})}}}\ I_{\gamma}^{UL}(> E^{th}_{\gamma}),
\label{eq:CRUL}
\end{equation}

\noindent
where $L_{pr}^{UL}$ is the upper limit on the proton UHECR luminosity, $I_{\gamma}^{UL}(> E^{th}_{\gamma})$ is the upper limit on the integral gamma-ray flux for a given confidence level and energy threshold, $K_{\gamma}$ is the number of gamma-rays generated from the cosmic-ray particles, $P_{\gamma}(E_{\gamma})$ is the energy distribution of the gamma-rays arriving on Earth, $E_{\gamma}$ is the energy of gamma-rays, $\langle E \rangle_{0}$ is the mean energy, $D_s$ is the comoving distance and $z_s$ is the redshift of the source.  Our model
can give a upper limits on the $L_{pr}^{UL}$ and
$\eta_{pr}^{UL}$ only in case radiation losses during acceleration of
UHECR are neglected as in the proposed model in \cite{supa,vitor}. The $L_{acc}^{Theory}$ is given by 

\begin{equation}
L_{acc}^{Theory} = \epsilon  \dot{M}c^2 = \frac{GM_{BH}\dot{M}}{6R} \; ,
\end{equation}

\noindent
where $R$ is the Kerr horizon and $\dot{M}$ is the Bondi, Kerr slim disk or Eddington models for accretion, and $M_{BH}$ is the mass of the central Seyfert black hole. 

\subsection{Estimating X-ray luminosity for massive Seyferts}

Active galactic nuclei (AGN), including quasars and Seyfert galaxies can emit a broad band spectral energy distribution (SED). These 
AGN emit from radio up to X-rays and sometimes gamma rays \cite{gu}. The X-rays luminosities are one of the evidence of nuclear activity and
hence are essential to study the accretion processes \cite{soria}. The bolometric luminosity of Seyferts represents the rate of
energy emitted by the accreting black hole. There is a dependence between the mass accretion rate and the efficiency of the acreting
matter \cite{veeresh}. We have estimated the X-rays luminosities of sources using the relation linearly proportional to the accretion rate,

\begin{equation}\label{equationlx}
L_{X}^{Theory} = L_{acc}^{Theory} = \xi \frac{GM_{BH}\dot{M}}{6R};
\end{equation}

\noindent
where $\xi$ is the efficiency factor with values between $\xi \approx
0.3$ for an accretion disk and $\xi \approx 10^{-11}$ for spherically
symmetry accretion \cite{jeremiah}. We consider $\xi \approx 0.03$ because the model of
a Seyfert accretion disk presents an inefficient accretion flow as
we have discussed in the previous section. We compare the observed X-ray luminosities of the seyferts with those expected from the
model accretion rate that we used here. The
results to $L_{x}^{Theory}$ in table \ref{tab:1} suggest that the X-ray
flux and the accretion model are consistent.

\subsection{Magnetic field and maximum energy extraction by massive Seyferts}\label{sec:mag}

Since the expected radiative energy transfer are not proeminently large to trigger UHECRs at the range established in the previous section (between the thresholds attached to Kerr slim disks and Bondi accretion mechanisms), now it will be considered a most promising mechanism for Seyferts, where a poloidal magnetic field attached to the slim disk and the central black hole can handle to extract energy from the disk or the black hole. 

In this case, the upper limit $B_{BH}$ of the magnetic field is \cite{zhang}
\begin{equation}
B_{BH} \sim 10^{8}\Biggl(\frac{M_{BH}}{M_{\odot}}\Biggl)^{-1/2}G,
\end{equation}
where the corresponding energy from the rotating Kerr BH/disk naturally should not exceed the Eddington luminosity \cite{carlos}. An useful method to obtain the relation between $B_{BH}$ and $M_{BH}$ was described in \cite{restrictions}. Using the Hillas condition with constraints and losses interaction losses and geometry of the source, the maximal energy is determined by condition
\begin{equation}\label{energy}
E_{max} \simeq 3.7\times 10^{19} eV \frac{A}{Z^{1/4}}\Biggl(\frac{M_{BH}}{10^{8}M_{\odot}}\Biggl)^{3/8}
\end{equation}
where $Z$ an $A$ are the atomic number and the mass of nuclei, respectively. The UHECR luminosity comes from the cosmic ray spectrum following the references
\cite{supa, vitor} and using a power law with an exponential cutoff of
$\alpha = 2.3$ and $E_{cut}$ given by equation \ref{energy}. 

Figure \ref{fig:fit} shows the correlation between the fraction ofluminosity of cosmic rays from the luminosity of the accreting black hole relative to the mass of the hole at the center of a massive Seyfert
for different models of accretion. The fraction decreases with increasing mass, the greater the mass the smaller the converted fraction luminosity of cosmic rays. Black hole masses were obtained using the method of Reines and Volonteri \cite{reines}, which estimates the correlation between the $M_{BH}$ and the total stellar masses in the
nearby universe ($z < 0.055$). The black hole mass is also related with
the velocity of the gas region indicating a relationship between mass
and black hole activity, \cite{wandel, jong} and the accretion can be
used to determine black hole mass \cite{greene}. The strong evidence between the black hole mass with the large-scale  properties  of  their  host  galaxies,  primarily the  bulge  component, was explicitly calculated, e.g., in \cite{magorrian,gebhardt,ferrarese,marconi,kormendy,mcconnell}.

The Fig. \ref{fig:emax} shows the relation (\ref{energy}) between
$E_{max}$ and $M_{BH}$ for iron, proton and carbon nuclei. Maximal
energy of protons for massive Seyfert were obtained and indicate that massive
Seyfert can accelerate protons to energies $\sim 10^{19.5}$ eV.

\section{Discussion}\label{sec:discussion}

In AGN and quasars, while it is considered that the strongest jets that accelerate particles result from the secular accumulation of magnetic flux \cite{sikora2007,marek,sikora2013}, moderate jet activity can also be triggered by fluctuations in the magnetic flux deposited by turbulent, hot inner regions of thin or slim accretion disks. These processes could be responsible for jet production in Seyferts and low-luminosity AGN \cite{marek}. We here consider the mechanisms by which a poloidal magnetic field threading the disk or the hole can manage to extract energy from the disk or the central BH. This energy can be extracted in the form of Poynting flux (i.e. purely electromagnetic energy) or in the form of a magnetically driven material wind. 

First of all, Blandford and Znajek \cite{blandford1977} noted that if the poloidal magnetic field trapped to the central black hole is comparable in strength to the poloidal field threading the inner parts of the accretion disk, then the BH contribution to the electromagnetic output is likely to be ignorable. To build up a poloidal field on the BH that considerably surpasses the poloidal field threading the inner disk two physical processes should occur: the disk must be able to transport poloidal field radially inwards and a field through the BH must be maintainable at the inner edge of the disk \cite{livio1999}.

\begin{figure}[h]
  \centering
  \includegraphics[width=12cm]{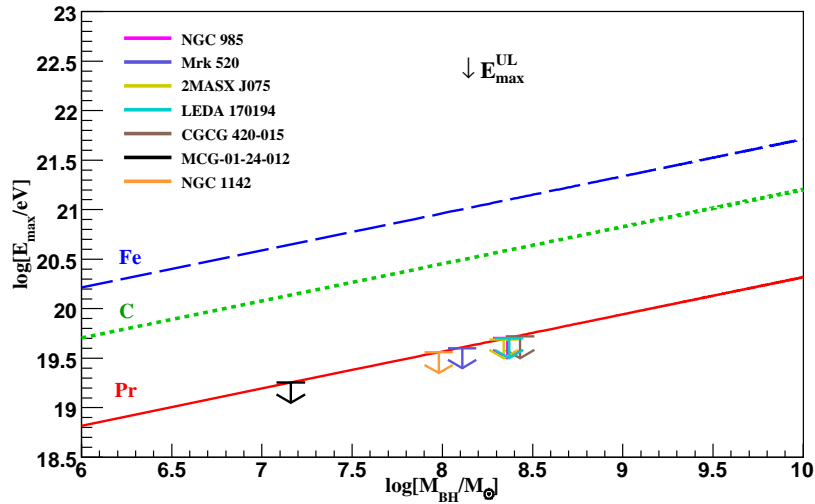}\\
\caption{Maximal energy for nuclei of iron, proton and carbon as a
  function of the $M_{BH}$. The arrows indicate the upper limit on the
  proton energy to Seyfert galaxies MCG-01-24-012, NGC 985 and NGC
  1142. The BH masses for the mentioned Seyferts come from \cite{vasu,franca}.}
\label{fig:emax}
\end{figure}

Sikora et al. (2013) \cite{sikora2013} suggested that in radio-loud AGN
the large magnetic fluxes were accumulated during a hot,
low-accretion-rate phase prior to the current, cold accretion
event. Regarding this, the nowadays cold and inneficient accretion flow
in Seyferts cannot accelerate particles by pure radiative mechanisms but
the past accumulation of magnetic flux is now operating the task at
least for the most massive Seyferts, as we can see in Figure
\ref{fig:emax}. Nevertheless, if the Seyfert is now sufficient massive,
we also can consider another mechanism to accellerate
particles. According to the magnetic flux paradigm, Seyferts galaxies
have very inefficient flux accumulation prior to the start of the
Seyfert activity phase. This inefficiency can be attributed to the lack
of a hot accretion pre-phase \cite{sikora2007,marek}. Nevertheless,
X-ray activity in Seyfert cores indicates a presence of a hot,
radiatively inefficient accretion zone extending out to some distance
from the BH (see Figure \ref{fig:sketch}). This indicates the presence
of thick flows that enable a significant poloidal magnetic field tot
impinge on the black hole and thus generate jets or winds.

Furthermore, we can model Seyfert disks as thin annuli and the efficient removal of angular momentum from a particular annulus leads to inward movement, outward bending of poloidal field lines, and consequently enhanced wind outflow, enhanced removal of angular momentum, and further inflow. As remarked by \cite{livio1999} in such case, if the central part of the AGN contains non-axisymmetric instabilities, as we imagine that happens in the case of Seyferts (X-ray core emission, thick of quasi-spherical central accretion, unstable jets), then it will possible to give rise to significant outward transport of angular momentum. In such case, a magnetically driven disk wind might be the main mechanism by which excess angular momentum is removed from disk material, and so might be the main mechanism that drives an accretion disk.

When poloidal magnetic winds cand trigger the accelleration engine of charged particles, as happens specially in the case of massive Seyferts, then such galaxies can be considered as Blandford-Znajek magnetically driven AGNs, with noticeable jets or outflows in the form of $\gamma$-rays or UHECRs. Indeed,  the detection of Narrow Line Seyfert 1 galaxies in the Fermi-LAT energy regime reveals essential hints to develop the understanding of jet formation, radio-loudness and particle acceleration at non-blazar cores \cite{rieger1,rieger2}.

Proton and heavy nuclei acceleration can also happens in the vicinity of the central BH horizon from the formation of vacuum gaps in the polar cap regions of a rotating BH surrounded by an accretion disk.
This paradigm, for a  reasonable range of black holes masses (as $10^6 M_{\odot}$-$10^{10} M_{\odot}$), works when the magnetic field is misaligned with the BH rotation axis, and after the charged particles penetrate the reconnection region they are leaked into the polar caps, being ejected with high energies. For example, in \cite{neronov} numerical modeling based in such mechanism led to the result that electromagnetic luminosity of the gap is of the order of the AGN bolometric luminosity, permiting, e.g., that local galaxies as Seyferts could be plausible sources of UHECRs, conclusion similar to the results calculated in the present paper.

\section{Concluding remarks}\label{sec:conclusion}

The present work raised an overview of the techniques employed, and the results obtained to classify massive Seyferts galaxies as possible
sources of cosmic rays. Firstly, Fig. 2 shows how Kerr black holes, using a axisymmetric accretion with a central spinning hole, according
to black masses, will produce UHECRs. A $M_{BH}^{-5.8}$ power law is
obtained, according to the black hole mass. The connection between
$L^{\mathrm{Theory}}_{CR}$ upper limits and the theoretical Seyfert
luminosity is explicitly investigated here by a conversion rate
$\eta_{pr}^{UL}$ calculated using the $E_{cut}$ from Hillas condition
(eq. $\ref{energy}$). The comparison among $L^{\mathrm{Theory}}_{CR}$,
the upper limit $L_{pr}$ and Seyfert theoretical luminosity offer limits
to the conversion fraction $\eta_{pr}^{UL}$ for each source. The high
resolution of Cherenkove Telescope Array (CTA) will allows the detection
of the emission from the active nuclei with energy range sensitivity of up to five orders of magnitud
and therefore the number of observed sources as black holes in the next
decades~ \cite{cta}. Here, the accretion mechanism has spherical symmetry (Bondi mechanism) and also axial symmetry (Kerr slim disk, Fig. 2), where it is assumed the general case of Kerr central black holes.

In second place, recent and ongoing efforts are rapidly expanding and revising the empirical black hole scaling relations, mainly between black hole masses and their correlation with the AGN host masses. Here we use this to some central black holes located in Seyferts, showing that massive Seyferts, chiefly, are the strong candidates to accelerate UHECRs (see Fig. 1 and 2). Mass conditions of Seyfert Nuclei in the close universe are sufficient to accelerate such particles, since central black holes present in such Seyferts only produce cosmic rays beyond the GZK limit for $M_{BH}> 10^7 M_\odot$ (Fig. 3) that are present in massive Seyferts \cite{kalm,ury,ammando,reines}.

\begin{acknowledgments}

The authors are very grateful to Vitor de Souza for helpful discussions and researchers of DEE-UFPR. This work was funded by CNPq under grant 458896/2013-6. 

\end{acknowledgments}

\end{document}